\documentclass[copyright,creativecommons]{eptcs}
\providecommand{\event}{EXPRESS'10} 
\usepackage{breakurl}             

\usepackage{epsfig}
\usepackage{latexsym}
\usepackage{graphicx}
\usepackage{color}
\usepackage{url}
\usepackage{stmaryrd}
\usepackage{amsmath}
\usepackage{tikz}

\newcommand{\post}[1]{\mbox{$#1^{\bullet}$}}
\newcommand{\pre}[1]{\mbox{$^{\bullet}#1$}}

\newcommand{\fine}{{\mbox{ }\nolinebreak\hfill{$\Box$}}}

\newcommand{\dec}{\mbox{$dec$}}
\newcommand{\sost}[2]{\mbox{$\{#1/#2\}$}}
\newcommand{\sostd}[4]{\mbox{$\{#1/#2,#3/#4\}$}}
\newcommand{\deriv}[1]{{\mbox{${\:\stackrel{#1}{\longrightarrow}\:}$}}}

\newcommand{\eqdef}{{\mbox{${\stackrel{def}{=}}$}}}
\newcommand{\spazio}{\hspace{2.4em}}
\newcommand{\bigfrac}[2]{
\renewcommand{\arraystretch}{1.5}
\begin{array}{c}#1\\
\hline
#2
\end{array}}
\newcommand{\restr}[1]{\mbox{$({\bf\nu} #1)$}}
\newcommand{\para}{\mbox{$\,|\,$}}

\newcommand{\parti}{\mbox{\raisebox{.4ex}{$\wp$}}}

\newcommand{\nil}{\mbox{\bf 0}}

\renewcommand{\mid}{\;\;\;\big|\;\;\;}

\newcommand{\nat}{{\rm I\!N}}

\newcommand{\IMG}{\mathit{IMG}}

\newcommand{\Sync}{\mathit{Sync}}

\newcommand{\inet}[1]{\ensuremath{INet(#1)}}

\newtheorem{adefinition}{Definition}
\newtheorem{atheorem}[adefinition]{Theorem}
\newtheorem{aproposition}[adefinition]{Proposition}
\newtheorem{acorollary}[adefinition]{Corollary}
\newtheorem{aexample}{\it Example}
\newenvironment{definition}{\begin{adefinition}}{\end{adefinition}}
\newenvironment{theorem}{\begin{atheorem} }{\end{atheorem}}
\newenvironment{proposition}{\begin{aproposition}}{\end{aproposition}}
\newenvironment{corollary}{\begin{acorollary} }{\end{acorollary}}
\newenvironment{example}{\begin{aexample} }{\end{aexample}}
\newcommand{\proof}{{\it Proof:}\ \ }

\title{A Process Calculus for Expressing Finite\\ Place/Transition Petri Nets}
\author{Roberto Gorrieri and Cristian Versari
\institute{Dipartimento di Scienze dell'Informazione, Universit\`a di
Bologna, \\Mura A. Zamboni, 7,
40127 Bologna, Italy}
\email{\{gorrieri, versari\}@cs.unibo.it}
}
\def\titlerunning{A Process Calculus for Expressing Finite P/T nets}
\def\authorrunning{R. Gorrieri \& C. Versari}

\begin{document}

\maketitle

\begin{abstract}
We introduce the process calculus Multi-CCS, which extends conservatively CCS with an operator of strong 
prefixing able to model atomic sequences of actions as well as multiparty synchronization. 
Multi-CCS is equipped with a labeled transition system semantics, which makes use of a minimal
structural congruence.
Multi-CCS is also equipped with an unsafe P/T Petri net semantics by means of a novel technique.
This is the first rich process calculus, including CCS
as a subcalculus, which receives a semantics in terms of unsafe, labeled P/T nets.
The main result of the paper is that a class of Multi-CCS processes, called {\em finite-net processes}, is able to represent
all finite (reduced) P/T nets. 
\end{abstract}

\section{Introduction}
Labeled transition systems with finitely many states and transitions can be expressed by the CCS \cite{Mil89} 
sub-calculus of {\em finite-state processes}, i.e., the sequential processes generated from
the empty process $\nil$, prefixing $\mu.p$, alternative composition $p_1+p_2$ and a finite number
of process  constants $C$,
each one equipped with a defining equation $C \eqdef p$. Intuitively, each state $s_i$ is modeled by
a constant $C_i$, whose defining equation contains one summand $a_j.C_j$ for each transition leaving state $s_i$
labeled by action $a_j$ and reaching the state $s_j$.
This celebrated result of Milner offers a process calculus to express, up to isomorphism, all finite-state labeled transition systems. The main advantage of this result is that $(i)$ finite-state lts's can be defined compositionally, and
$(ii)$ behavioral equivalences over finite-state lts's can be axiomatized \cite{Mil89b}.

This paper addresses the same language expressibility problem for finite labeled Place/Transition Petri nets
without capacity bounds on places. 
We single out a fragment of an extension of CCS, called Multi-CCS,
such that not only all processes of this fragment generate finite P/T nets, but also for any finite (reduced) P/T net
we can find a term of the calculus that generates it. This solves the open problem of 
providing a process calculus for general Petri nets. 
and opens interesting possibilities of cross-fertilization between the areas of Petri nets and process calculi.
In particular, it is now possible, on the one hand, $(i)$  to define any finite P/T net compositionally and $(ii)$ to start the investigation
of axiomatization for behavioral equivalences over such a large class of nets; on the other hand, it is now possible
$(iii)$ to reuse all the techniques and decidability results available for P/T nets also for (this fragment of) Multi-CCS,
as well as $(iv)$ define non-interleaving semantics, typical of Petri nets, also for Multi-CCS.

We equip Multi-CCS with an operational net semantics that takes inspiration from Goltz's idea of using unsafe, 
labeled P/T nets \cite{Ulla1,Ulla,GMa} for a CCS subcalculus without restriction, 
and Busi \& Gorrieri net semantics for $\pi$-calculus \cite{BG09}, where however
inhibitor arcs are used to model restriction.  The extension of the approach to restriction and strong prefixing is not
trivial and passes through the introduction of an auxiliary set of {\em restricted} actions and
the definition of a suitable notion of syntactic substitution.
We prove a soundness result, i.e.,  $p$ and $Net(p)$ are strongly bisimilar, where the net $Net(p)$ is the
subnet reachable from the marking associated to process $p$. 

The Multi-CCS sub-calculus of {\em finite-net processes} is generated as follows:\\[-.6cm]
\begin{eqnarray*}
s & ::= & \nil \mid \mu.t \mid \underline\mu.t \mid s+s \\[-.1cm]
t & ::= & s \mid t \para t \mid C \\[-.1cm]
p & ::= &   t  \mid \restr{a}p \mid p \para p\\[-.7cm]
\end{eqnarray*}
where the operator $\underline\mu.t$, called {\em strong prefixing} (in opposition to normal prefixing),  expresses that action $\mu$ is the initial 
part of an atomic sequence of actions that continues with $t$. This operator, introduced in \cite{GMM}, is also at
the base of multiparty synchronization, obtained as an atomic sequence of binary CCS-like synchronizations.
As a constant $C \eqdef t$, we have that parallel composition $\para$ may occur inside the body $t$ 
of recursively defined constants; hence, finite-net processes are infinite-state processes.
On the contrary, restriction $\restr{a}$ is not allowed in the body of recursively defined constants. 
We also require that the alternative
composition $+$ is guarded, i.e., all summands are sequential. Finally, constants are
assumed to be {\em guarded}, i.e., in any defining equation each occurs inside a {\em normally prefixed} subprocess
$\mu.t$.

We prove that the operational net semantics associates a finite P/T net $Net(p)$ to any finite-net process $p$.
Conversely, we also prove that for any finite reduced P/T net $N$, we can find a finite-net process $p_N$ such that $Net(p_N)$
and $N$ are isomorphic. The construction of the finite-net process $p_N$ from a net $N$ associates to each place $s_i$
of the net a process constant $C_i$, whose defining equation contains one summand for each transition for 
which place $s_i$ is an input; moreover, as multiparty synchronization is implemented as an atomic sequence of binary
synchronizations, for each transition there is the need to elect a leader among its places in the preset that 
coordinates the actual multiparty synchronization. Some examples are presented to illustrate the approach.

The paper is organized as follows. Section 2 contains some basic background. Section 3 introduces the process 
calculus Multi-CCS, together with some examples (dining philosophers and concurrent readers/writers). 
Section 4 defines the operational net semantics for Multi-CCS. Section 5 provides
the soundness theorem ($p$ and $Net(p)$ are bisimilar) and the finiteness theorem (for any finite-net process $p$,
$Net(p)$ is finite). Section 6 proves the language expressibility theorem (for any finite reduced P/T net $N$ there exists 
a finite-net process $p_N$ such that $N$ is isomorphic to  $Net(p_N)$). Finally, some conclusions are drawn in Section 7.

\section{Background}%

\subsection{Labeled transition systems and bisimulation}%

\begin{definition}
A labeled transition system is a triple $TS = (St, A,$ $ \rightarrow)$ where
$St$ is the set of states,
$A$ is the set of labels,
$\rightarrow\subseteq St\times A\times St$ is the transition relation.
In the following  $s\deriv{a}s'$ denotes $(s,a,s')\in \rightarrow$.
A {\em rooted} transition system is a pair $(TS, s_0)$ where $TS = (St,
A, $ $\rightarrow)$ is a transition system and $s_0\in St$ is the {\em initial
state}.
\end{definition}

\begin{definition}
A {\em bisimulation} between $TS_1$ and $TS_2$ is a relation
$R\subseteq (St_1\times St_2)$ such that  if $(s_1, s_2) \in R$
then for all $a \in (A_1 \cup A_2)$\\[-.6cm]
\begin{itemize}
\item $\forall s'_1$ such that  $s_1\deriv{a}s_1'$, $\exists s'_2$ such that $s_2\deriv{a}s_2'$ and $(s_1',s_2') \in R$
\item $\forall s'_2$ such that $s_2\deriv{a}s_2'$, $\exists s'_1$ such that $s_1\deriv{a}s_1'$ and $(s_1',s_2') \in R$.\\[-.6cm]
\end{itemize}
If $TS_1 = TS_2$ we say that $R$ is a bisimulation  on $TS_1$. Two states $s$ and $s'$ are bisimilar,  
$s \sim s'$, if there exists a bisimulation $R$ such that $(s, s') \in R$.
\end{definition}

\subsection{Place/Transition Petri nets}


\begin{definition}
Let $\nat$ be the set of natural numbers. 
Given a set $S$, a {\em finite multiset} over $S$ is a function $m: S \rightarrow\nat$ such 
that the set $dom(m) = \{ s \in S \, | \,m(s) \neq 0\}$ is finite. The {\em multiplicity} of 
 $s$ in $m$ is given by the number $m(s)$. 
The set of all finite multisets 
over $S$,  ${\mathcal M}_{fin}(S)$, is ranged over by $m$. 
$\parti_{fin}(S)$ is the set of all finite sets over $S$.
We write $m \subseteq m'$ if $m(s) \leq m'(s)$ for all $s \in S$. The operator $\oplus$ 
denotes {\em multiset union}: $(m \oplus m')(s)$ $ =m(s) + m'(s)$.
The operator $\setminus$ 
denotes (limited) {\em multiset difference}: $(m \setminus m')(s) =$ if $m(s)>m'(s)$ then  $m(s)-m'(s)$ else $0$.
The {\em scalar product} of a number $j$ with $m$ is $(j \cdot m)(s) = j \cdot (m(s))$.
A finite multiset $m$ over $S = \{s_1, s_2, \ldots  \}$ can be also represented as
$k_1s_{i_1} \oplus k_2s_{i_2} \oplus \ldots \oplus k_n s_{i_n}$, where $dom(m) = \{s_{i_1}, \ldots s_{i_n} \}$ and
$k_j = m(s_{i_j})$ for $j= 1, \ldots, n$.
\end{definition}

\begin{definition}
A labeled P/T Petri net is a tuple $N = (S, A, T)$, where
$S$ is the set of {\em places},
$A$ is a set of labels and
$T \subseteq {\mathcal M}_{fin}(S) \times A \times {\mathcal M}_{fin}(S)$ is the set of transitions.
A P/T net is {\em finite} if both $S$ and $T$ are finite.
A finite multiset over $S$  is called a {\em marking}. Given a marking $m$ and a place $s$, 
we say that the place $s$ contains $m(s)$ {\em tokens}. 
Given a transition $t = (m, a, m')$,
we use the notation  $\pre t$ to denote its {\em preset} $m$, $\post t$ for its {\em postset} $m'$
and $l(t)$ for its label $a$. Hence, transition $t$ can be also represented as $\pre t \deriv{l(t)} \post t$.
\end{definition}


\begin{definition}
Given a labeled P/T net  $N = (S, A, T)$, we say that
a transition $t $ is {\em enabled} at marking $m$, written as $m[t\rangle$, if $\pre t \subseteq m$. 
The execution of  $t$ enabled at $m$ produces the marking $m' = (m \setminus  \pre t) \oplus \post t$. 
This is written as $m[t\rangle m'$. 

A {\em P/T system} is a tuple $N(m_0) = (S, A, T, m_{0})$, where $(S,A, T)$ is a P/T net and $m_{0}$ is a  finite
multiset over $S$, called the {\em initial marking}. 
The set of markings {\em reachable} from $m$, denoted $[m\rangle$, is defined as the least set  such that
$m\in [m\rangle$ and
if $m_1\in [m\rangle$ and, for some transition $t\in T$, $m_1[t\rangle m_2$, then $m_2\in [m\rangle$.
We say that $m$ is reachable if $m$ is reachable from the initial marking $m_{0}$.
A P/T system is said to be {\em safe} if any place contains at most one 
token in any reachable marking, i.e. $m(s)\leq 1$ for all $s\in S$ and for all $m\in[m_{0}\rangle$.
\end{definition}

\begin{definition}\label{reduced}
A P/T system $N(m_0) = (S, A, T, m_{0})$ is {\em reduced} if
 $\forall s \in S \; \exists m \in [m_0\rangle$
such that $m(s) \geq 1$, and 
$\forall t \in T \; \pre t \neq \emptyset \wedge  \exists m \in [m_0\rangle$ such that $m[t\rangle$.
\end{definition}


\begin{definition}
The {\em interleaving marking graph} of $N(m_0)$ is the lts
$\IMG(N(m_0))$ $ = ([m_{0}\rangle, A, \rightarrow, m_{0})$, where $m_0$ is the initial state 
and the transition relation 
is defined by $m\deriv{l(t)}m'$ iff there exists a transition $t\in T$ such that $m[t\rangle m'$. 
The P/T systems $N_1(m_{1})$ and $N_2(m_{2})$ are {\em interleaving bisimilar} ($N_1\sim N_2$) iff there exists 
a strong bisimulation  relating the initial states of  $\IMG(N_{1}(m_1))$ and $\IMG(N_{2}(m_2))$.
\end{definition}

\begin{definition}
Given two P/T net systems $N_1( m_{0_1})$ and $N_2(m_{0_2})$, we say that $N_1$ and $N_2$ are {\em isomorphic}
if there exists a bijection $f: S_1 \rightarrow S_2$, homomorphically extended to markings, such that
$f(m_{0_1}) = m_{0_2}$ and
$(m, a, m') \in T_1$ iff $(f(m), a, f(m')) \in T_2$.
\end{definition}

\section{Multiparty synchronization in CCS} \label{basic}

In this section we present  Multi-CCS, obtained as a variation over A$^2$CCS  \cite{GMM,GM90}; the
main differences are that in Multi-CCS the parallel operator is associative, and the synchronization
relation on sequences is less verbose. Then,  two case studies are presented.


\subsection{Multi-CCS}

Let ${\mathcal L}$ be a denumerable set of channel names, ranged over by 
$a,b,\ldots$. Let  $\overline{\mathcal L}$ the set of co-names, ranged over by 
$\overline{a}, \overline{b},\ldots$. The set  ${\mathcal L}\cup\overline{\mathcal L}$, ranged over by $\alpha, \beta, \ldots$,
is the set of visible actions. With $\overline \alpha$ we mean the complement of $\alpha$, assuming that $\overline{\overline \alpha} = \alpha$.
Let $Act = {\mathcal L} \cup \overline{\mathcal L} \cup\{\tau\}$,
such that $\tau\not\in{\mathcal L}\cup\overline{\mathcal L}$, be the set of actions, ranged over by 
$\mu$. Action $\tau$ denotes an invisible, internal activity. 
Let ${\mathcal C}$ be a denumerable set of process constants, disjoint from 
$Act$, ranged over by $A, B, C,\ldots$.
The process terms are generated from actions and constants by:\\[-.4cm]
\[p ::= \nil \mid \mu.q \mid  \underline{\mu}.q \mid p+p \; \; \mbox{  {\em sequential processes}}\]\\[-.9cm]
\[q ::= p \mid q\para q \mid \restr{a}q \mid C  \; \; \mbox{  {\em processes}}\]\\[-.5cm]
where $\nil$ is the terminated process, $\mu.q$ is a normally prefixed process where action $\mu$ (that
can be either an input $a$, an output  $\bar{a}$ or a silent move $\tau$) is first performed and then $q$ is ready, 
$ \underline{\mu}.q$ is a  strongly prefixed process where $\mu$ is the first action of a transaction that continues with $q$
(provided that $q$ can complete the transaction), 
$p + p'$ is the sequential process obtained by the alternative composition of sequential processes $p$ and $p'$, 
$q \para q'$ is the parallel composition of $q$ and $q'$, $\restr{a}q$ is process $q$ where the (input) name $a$ is made private (restriction),
$C$ is a process constant, equipped with a defining equation $C \eqdef q$.

The set ${\mathcal P}$ of {\em processes} contains those terms which are, w.r.t. process constants they use,
{\em closed} (all the constants possess a defining equation) and {\em guarded} (for any defining equation $C \eqdef q$,
any occurrence of $C$ in $q$ is  within a {\em normally prefixed} subprocess $\mu.q'$ of $q$). 
With abuse of notation, ${\mathcal P}$ will be ranged over by $p, q,\ldots$.  
${\mathcal P}_{seq}$ is the set of  {\em sequential processes}.

The  operational semantics for Multi-CCS is given by the labelled 
transition system $({\mathcal P},{\mathcal A}, \deriv{ })$, where the states are the processes
in ${\mathcal P}$, ${\mathcal A} = Act^*$ 
is the set of labels (ranged over by $\sigma$),
and $\deriv{ } \subseteq {\mathcal P}\times{\mathcal A}\times{\mathcal P}$ is the minimal 
transition relation generated by the rules listed in Table~\ref{rules}.

\begin{table}[t]
{\renewcommand{\arraystretch}{1.8}
\hrulefill\\[-.8cm]

\begin{center}
$\begin{array}{lcllcl}
\mbox{(Pref)}  & \mu.p\deriv{\mu}p & & \; \; \;
\mbox{(S-pref)}  & \bigfrac{p\deriv{\sigma}p'}{\underline{\mu}.p\deriv{\mu\sigma}p'} & \\
\mbox{(Sum)}  & \bigfrac{p\deriv{\sigma}p'}{p+q\deriv{\sigma}p'} &  & \; \; \;
\mbox{(Com)}  & \bigfrac{p\deriv{\sigma_1} p'\spazio q\deriv{\sigma_2}q'}{p
\para q \deriv{\sigma}p'\para q'} & \Sync(\sigma_1,  \sigma_2, \sigma) \\
\mbox{(Par)}  & \bigfrac{p\deriv{\sigma}p'}{p\para q\deriv{\sigma}p'\para q} & & \; \; \;
\mbox{(Res)}  & \bigfrac{p\deriv{\sigma}p'}{\restr{a}p\deriv{\sigma}
\restr{a}p'} & a, \bar{a} \not\in n(\sigma) \\
\mbox{(Cong)} & \bigfrac{p \equiv p' \deriv{\sigma}q' \equiv q}{p\deriv{\sigma}q}  & & \; \; \;
\mbox{(Cons)} & \bigfrac{p\deriv{\sigma}p'}{C\deriv{\sigma}p'}& C \eqdef p  \\
\end{array}$

\hrulefill
\end{center}}
\caption{Operational semantics (symmetric rules for (Sum) and (Par) omitted)}\label{rules}
\end{table}

\begin{table}[t]
\hrulefill\\[-.8cm]
\begin{center}
$\begin{array}{lll}
Sync(\alpha, \bar{\alpha}, \tau) &  \; \; \bigfrac{\sigma \neq \epsilon}{Sync(\alpha\sigma, \bar{\alpha}, \sigma)} &
\; \; \bigfrac{\sigma \neq \epsilon}{Sync(\alpha, \bar{\alpha}\sigma, \sigma)} \\
\bigfrac{Sync(\sigma_1, \sigma_2, \sigma)}{Sync(\alpha\sigma_1, \bar{\alpha}\sigma_2, \sigma)} &
\; \;  \bigfrac{Sync(\sigma_1, \sigma_2, \sigma)}{Sync(\alpha\sigma_1, \sigma_2, \alpha\sigma)} & 
\; \; \bigfrac{Sync(\sigma_1, \sigma_2, \sigma)}{Sync(\sigma_1, \alpha\sigma_2, \alpha\sigma)} \\
\bigfrac{Sync(\sigma_1, \sigma_2, \sigma)}{Sync(\tau\sigma_1, \sigma_2, \sigma)} &  
\; \; \bigfrac{Sync(\sigma_1, \sigma_2, \sigma)}{Sync(\sigma_1, \tau\sigma_2, \sigma)} \\[-.3cm]
\end{array}$
\end{center}

\hrulefill \\[-.3cm]
\caption{Synchronization relation}\label{sync}
\end{table}

We briefly comment on the rules that are less standard. 
Rule (S-pref) allows for the creation of
transitions labeled by non-empty sequences of actions. In order for $\underline{\mu}.q$ to make a move, it is
necessary that $q$ can perform a transition, i.e., the rest of the transaction. Hence, $\underline \mu.\nil$ 
cannot perform any action. If a transition is labeled by $\sigma = \mu_1 \ldots \mu_n$, then all the
actions $\mu_1 \ldots \mu_{n-1}$ are due to strong prefixes, while $\mu_n$ to a
normal prefix.
Rule (Com) has a side-condition on the possible
synchronizability of sequences $\sigma_1$ and $\sigma_2$. $\Sync(\sigma_1, \sigma_2, \sigma)$ holds
if $\sigma$ is obtained from an interleaving (possibly with synchronizations) of $\sigma_1$ and $\sigma_2$, 
where the last action of one of the two sequences is to be synchronized, hence reflecting that the 
subtransaction that ends first signals this fact (i.e., {\em commits}) to the other subtransaction. Relation $\Sync$ is defined 
by the inductive rules of Table~\ref{sync}.  
Rule (Res) requires that no action in $\sigma$ can be $a$ or $\bar{a}$.
$n(\sigma)$ denotes the set of all actions occurring in $\sigma$.
Rule (Cong) makes use of a structural congruence $\equiv$ on process terms induced by the 
following three equations:

$
(p \para q) \para r \; = \; p \para (q \para r)
$

$
\restr{a} (p \para q) = p \para \restr{a}q \; \; \mbox{ if $a$ is not free in $p$}.
$

$
\restr{a} p =  \restr{b} (p\sost{b}{a}) \; \; \mbox{ if $b$ is not free in $p$}.
$

\noindent
The first equation is for associativity of the parallel operator; the second one allows for enlargement of the scope of restriction;
the last equation is the so-called law of {\em alpha-conversion}, which makes use of syntactic substitution.\footnote{In this paper we use a slightly different definition of syntactic substitution in that 
$(\restr{a}q)\sost{b}{a} = \restr{b}q\sost{b}{a}$ if $b$ is not free in $q$, so that also the bound name $a$ is 
converted. This is necessary in the net semantics, in order to be sure that a substitution $\sost{b}{a}$ will be eventually applied
to any inner constant $C$ (defined as $p$) in $q$; the result of $C\sost{b}{a}$ is a new constant 
$C_{\sost{b}{a}} \eqdef p\sost{b}{a}$. See Example \ref{counter} for an application of this idea.}
Rule (Cong) enlarges the set of transitions derivable from $p$, as the following example shows. 
Also, it is necessary to ensure validity of Proposition \ref{altravia}.

\begin{example}\label{multi-synch}{\bf (Multi-party synchronization)} Assume three processes want to synchronize. 
This can be expressed in Multi-CCS. E.g., consider processes $p = \underline{a}.a.p'$,
$q = \bar{a}.q'$ and $r = \bar{a}.r'$ and the whole system $P = \restr{a}((p \para q) \para r)$. It is easy to see
that $P \deriv{\tau} \restr{a}((p' \para q') \para r')$ (and this can be proved in two ways), so the three processes 
have synchronized in one single atomic transition.
It is interesting to observe that $P' = \restr{a}(p \para (q \para r))$ could not
perform the multiway synchronization if  rule (Cong) were not allowed.
\end{example}

\begin{example}\label{guarded-rec}{\bf (Guardedness)} We assume that each process constant in a defining equation occurs
inside a normally prefixed subprocess $\mu.q$. This will prevent infinitely branching sequential processes.
E.g, consider the non legal process $A  \eqdef  \underline{a}.A + b.\nil$.
According to the operational rules, $A$ has infinitely many transitions leading to $\nil$, each of the form $a^nb$, for 
$n = 0, 1, ...$. 
\end{example}

Two terms $p$ and $q$ are {\em interleaving bisimilar}, written
$p\sim q$, if there exists a bisimulation $R$ such that $(p,q)\in R$.
Observe that $\restr{a}\restr{b}p \; \sim \;  \restr{b}\restr{a}p$,
which allows for a simplification in the notation that we usually adopt, namely restriction on 
a set of names, e.g., $\restr{a,b}p$.\\[-.6cm]

\subsection{Case studies}

\begin{example}\label{ex-dining}
{\bf(Dining Philosophers)} This famous problem, defined by Dijkstra in \cite{Dij71}, can be solved in Multi-CCS.
Five philosophers seat at a round table, with a private plate and where each of the five forks is shared by two neighbors.
Philosophers can think and eat; in order to eat, a philosopher has to acquire both forks that he shares with his neighbors,
starting from the fork at his left and then the one at his right. All philosophers behave the same, so the problem is intrinsically symmetric.
Clearly a na\"ive solution would cause deadlock exactly when all five philosophers take the fork at their left at the same time 
and are waiting for the fork at their right.
A simple solution is to force atomicity on the acquisition of the two forks. 
In order to have a small net model, we consider the case of two philosophers only.
The forks can be defined by the constants $fork_i$: \\[-.7cm]
\begin{eqnarray*}
fork_i \eqdef  \overline{up_i}.\overline{dn_i}.fork_i \; \;  \mbox{   for  }i = 0, 1  \\[-.7cm]
\end{eqnarray*}
The two philosophers can be described as \\[-.6cm]
\begin{eqnarray*}
phil_i \eqdef think.phil_i + \underline{up_i}.up_{i+1}.eat.\underline{dn_i}.dn_{i+1}.phil_i  \; \;  \mbox{   for  }i = 0, 1 \\[-.7cm]
\end{eqnarray*}
where $i+1$ is computed modulo $2$ and  the atomic sequence $up_i up_{i+1}$ ensures the atomic 
acquisition of the two forks. 
The whole system is  \\[-.7cm]
\begin{eqnarray*}
DF \;  \eqdef \;  \restr{L}(((phil_0\para phil_1) \para fork_0) \para fork_1) \\[-.7cm]
\end{eqnarray*}
where $L = \{up_0, up_1, dn_0, dn_1\}$. 
Note that the operational semantics generates a finite-state 
lts for $DF$.
\end{example}

\begin{example}\label{ex-conc-rw}
{\bf(Concurrent readers and writers)} There are several variants of this problem, defined in \cite{CHP71}, 
which can be solved in Multi-CCS. Processes are of two types: reader processes and writer processes. All processes
share a common file; so, each writer process must exclude all the other writers and all the readers while writing on the file,
while multiple reader processes can access the shared file simultaneously. Assume to have $n$ readers, $m$ writers and 
that at most $k \leq n$ readers can read simultaneously. A writer must prevent all the $k$ possible concurrent reading operations. 
A simple solution is to force atomicity on the acquisition of the $k$ locks so that either all are taken or none. 
To make the presentation simple, assume that $n = 4, k=3, m=2$.
Each reader process $R$, each lock process $L$, each writer $W$ can be represented as follows, where
action $l$ stands for $lock$ and $u$ for $unlock$ :

$
\begin{array}{lcllcllcl}
R &  \eqdef &  l.read.u.R & \; \; \;
L &  \eqdef & \overline{l}.\overline{u}.L & \; \; \;
W &  \eqdef &  \underline{l}.\underline{l}.l.write.\underline{u}.\underline{u}.u.W \\
\end{array}
$

$
\begin{array}{lcl}
Sys & \eqdef & \restr{l,u}((((((R \para R) \para (R \para R)) \para (W \para W)) \para L) \para L) \para L)\\
\end{array}
$

\noindent
It is easy to see that the labeled transition system for $Sys$ is finite-state. 
\end{example}

\section{Operational Net Semantics}\label{bg}

In this section we first describe a technique for building a P/T net for the whole Multi-CCS,
starting from a description of its places and of its net transitions. 
The resulting net $N_{MCCS} = (S_{MCCS}, {\mathcal A}, T_{MCCS})$ is such that,
for any $p \in {\mathcal P}$, the net system $N_{MCCS} (\dec(p))$ reachable from the intial
marking $\dec(p)$ is a reduced P/T net. 

\subsection{Places and markings}
The Multi-CCS processes are built upon the denumerable set  ${\mathcal L}\cup\overline{\mathcal L}$, ranged over by $\alpha$,  of visible actions. We assume to have another denumerable set ${\mathcal N}\cup\overline{\mathcal N}$ ranged over by $\delta$, of auxiliary {\em restricted} actions. The set of all actions $Act' = {\mathcal L}\cup\overline{\mathcal L} \cup {\mathcal N}\cup\overline{\mathcal N} \cup \{\tau\}$, ranged over by $\mu$ with abuse of notation, 
is used to build the enlarged set of processes we denote 
with ${\mathcal P}^{\mathcal N}$.

The infinite set of places, ranged over by $s$ (possibly indexed), is 
$S_{MCCS} = {\mathcal P}^{\mathcal N}_{seq}$, i.e., the set of all sequential processes over $Act'$.

\begin{table}[t]

{\renewcommand{\arraystretch}{2.0}
\hrulefill\\[-.6cm]
\begin{center}
$\begin{array}{lllll}
\dec(\nil) = \emptyset & 
\; \dec(\mu.p) = \{ \mu.p\}  &  \dec(\underline\mu.q) = \{\underline\mu.q\}\\
\dec(p+p') =   \{p+p'\}  & 
\; \dec(\restr{a}q) = \dec (q\sost{a'}{a}) &\; \;  a' \in {\mathcal N} \mbox{ is a new restricted action} \\
\dec(q \para q') = \dec(q) \oplus dec(q') & 
\; \dec (C) = \dec(p)  \; \; \mbox{ if }C \eqdef p\\[-.1cm]
\end{array}$
\end{center}}
\hrulefill \\[-.2cm]

\caption{Decomposition function} \protect\label{deca3}
\end{table}

Function $\dec: {\mathcal P}^{\mathcal N} \rightarrow {\mathcal M}_{fin}(S_{MCCS})$ (see Table~\ref{deca3})
defines the decomposition of processes into markings. Agent $\nil$ generates no places. The decomposition of
a sequential process $p$ produces one place with name $p$. This is the case
of $\mu.p$, $\underline\mu.p$ and $p+p'$.
Parallel composition is interpreted as multiset union;  the decomposition of, e.g., $a.\nil \para a.\nil$ 
produces the marking $a.\nil \oplus a.\nil = 2a.\nil$.
The decomposition of
a restricted process $\restr{a}q$ generates  the multiset obtained from the 
decomposition of $q$ where the new restricted name $a' \in {\mathcal N}$ is substituted for the
bound name $a$. 
Finally, a process constant is first unwound once (according to its defining equation) 
and then decomposed.

 It is possible to prove that the decomposition function $\dec$ is well-defined by induction on a 
 suitably defined notion of complexity of terms (following \cite{Old} page 52). 
Guardedness (even w.r.t. any kind of prefix) of constants is 
essential to prove the following obvious fact. 

\begin{proposition}
For any process $p \in {\mathcal P}^{\mathcal N}$, $\dec(p)$ is a finite multiset of places.
\fine
\end{proposition}

Note that $\dec$ is not injective; e.g., $\dec(a.\nil \para b.\nil) = \dec(b.\nil \para a.\nil)$.

Note that a fresh restricted name $a'$ is to be generated for each
of the $\dec$ applications on the right-hand-side of the transition schemata
we will describe in the next section. So in a recursive term, e.g., 
$A = {\restr{a}(a.A \para b.A)}$,
there may be the need for an unbounded number of fresh  names.

\subsection{Net transitions}

Let $\rightarrow \subseteq  {\mathcal M}_{fin}(S_{MCCS}) \times {\mathcal B} \times {\mathcal M}_{fin}(S_{MCCS})$,
where $ {\mathcal B}  = Act'^{*}$, be the least set of transitions generated by the rules in Table~\ref{netrulesa3}. 

Let $H, K$, possibly indexed, range over ${\mathcal M}_{fin}(S_{MCCS})$.
In a transition $H \deriv \sigma K$, $H$ is the multiset of tokens to be consumed,
$\sigma$ is the label of
the transition and $K$ is the multiset of tokens to be produced.

Let us comment the rules. 
Axiom (pref) states that if one token is present in $\{\mu.q\}$ then a $\mu$-labeled transition is derivable,
producing the tokens specified by $\dec(q)$. This holds for any $\mu$, i.e., for the invisible action $\tau$,
for any visible action $\alpha$ as well as for any restricted action $\delta$.
Transition labeled by restricted actions should not be taken in the resulting net, as we restrict ourselves to transitions labeled by sequence on visible actions only (and $\tau$). However, they are useful in producing normal synchronization, as two complementary restricted actions can produce a $\tau$-labeled transition. 
Rule (s-pref) requires that the premise transition $H \deriv{\sigma} H'$ is derivable 
by the rules, where $H$ is a submultiset of $\dec(q)$.
Rule (sum) is as expected.
Finally, rule (com) explains how synchronization takes place: it is needed that $H$ and $K$ perform
synchronizable sequences $\sigma_1$ and $\sigma_2$, producing $\sigma$; here we assume that 
$Sync$ has been extended also to restricted actions in the obvious way.

Note that transitions can be labeled also by restriction actions, while we are interested only in transitions that 
are labeled on ${\mathcal A} = Act^{*}$. 
Hence, the P/T net for Multi-CCS is the 
triple $N_{MCCS} = (S_{MCCS}, {\mathcal A},$ $ T_{MCCS})$, where the infinite set 
$T_{MCCS} = \{ (H, \sigma, K) \mid H \deriv{\sigma}K \wedge \sigma \in {\mathcal A} \}$  is obtained by filtering out 
those transitions where no restriction name $\delta$ occurs in $\sigma$.

\begin{table}
\hrulefill\\[-.6cm]
{\renewcommand{\arraystretch}{2.2}
\begin{center}
$\begin{array}{llll}
\mbox{(pref)} & \{\mu.q\} \deriv{\mu} \dec(q)  &
\; \; \mbox{(sum)} & \bigfrac{\{p\} \deriv{\sigma}H}{ \{p+p'\} \deriv{\sigma}H} \\
\end{array}$

$\begin{array}{lcl}
\mbox{(s-pref)} & \bigfrac{H \deriv{\sigma}H' }{ \{\underline\mu.q\} \deriv{\mu\sigma}H'\oplus K} &  H\oplus K = \dec(q)\\
\mbox{(com)} & \bigfrac{H \deriv{\sigma_1}H' \; K  \deriv{\sigma_2}K' }{ 
H \oplus K  \deriv{\sigma} H' \oplus K'} &  Sync(\sigma_1, \sigma_2, \sigma)\\
\end{array}$

\hrulefill
\caption{Rules for net transitions (symmetric rule for (sum) omitted).} 
\protect\label{netrulesa3}
\end{center}}
\end{table}

\begin{proposition}\label{decprop}
Let $t = H \deriv{\sigma}H'$ be a transition. Let $p$ be such that $\dec(p) = H \oplus K $
and let $t$ be enabled at $\dec(p)$. Then  $H' \oplus K = \dec(p')$ for some $p'$.

\proof
By induction on the definition of $\dec(p)$ and then on the proof of $t$ .
\fine
\end{proposition}

Given a process $p$, the P/T system associated to $p$ is the subnet of 
$N_{MCCS}$ reachable from the initial marking $\dec(p)$. We indicate with $Net(p)$
such a subnet.

\begin{definition}
Let $p$ be a process. 
The P/T system associated to $p$ is $Net(p) = (S_p, A_p, T_p, m_0)$, where $m_0 = \dec(p)$ and \\[-.8cm]
\begin{eqnarray*}
S_p & = & \{s\in S_{MCCS} \mid \exists m\in[m_0\rangle(m(s)>0)\}\\
T_p &=& \{ t \in T_{MCCS}  \mid \exists m\in[m_0\rangle \mbox{ s.t. } m[t\rangle \}\\
A_p &=& \{\sigma \in {\mathcal A} \mid \exists t \in T_p, \sigma = l(t))\}\\[-.7cm]
\end{eqnarray*}
\end{definition}

The definition above suggests a way of
generating $Net(p)$ with an algorithm in least-fixpoint style. Start by $\dec(p)$ 
and then apply the rules in Table \ref{netrulesa3} in order to produce the set of transitions
(labeled on ${\mathcal A}$) executable from $\dec(p)$ in one step. 
This will also produce possible new places
to be added to the current set of places. Then repeat until no new 
places are added and no new transitions are derivable; hence, this algorithm ends only for finite nets.

The following facts are obvious by construction:

\begin{proposition}
For any $p \in {\mathcal P}$, \\[-.5cm]
\begin{itemize}
\item $Net(p)$ is a reduced (see Definition \ref{reduced}) P/T net.\\[-.5cm]
\item $Net(p) \sim  N_{MCCS}(\dec(p))$.\\[-.6cm]
\end{itemize}
\end{proposition}

\subsection{Case Studies}

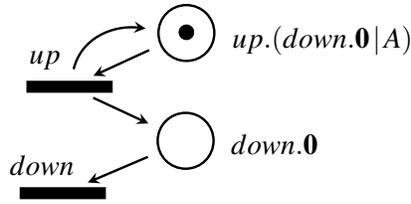
\begin{figure}\centering
\ifx\du\undefined
  \newlength{\du}
\fi
\setlength{\du}{3\unitlength}
\begin{tikzpicture}
\pgftransformxscale{1.000000}
\pgftransformyscale{-1.000000}
\definecolor{dialinecolor}{rgb}{0.000000, 0.000000, 0.000000}
\pgfsetstrokecolor{dialinecolor}
\definecolor{dialinecolor}{rgb}{1.000000, 1.000000, 1.000000}
\pgfsetfillcolor{dialinecolor}
\definecolor{dialinecolor}{rgb}{0.000000, 0.000000, 0.000000}
\pgfsetstrokecolor{dialinecolor}
\node[anchor=west] at (39.655000\du,9.987500\du){$up.(down.\mathbf{0} \para A)$};
\definecolor{dialinecolor}{rgb}{0.000000, 0.000000, 0.000000}
\pgfsetstrokecolor{dialinecolor}
\node[anchor=west] at (39.502100\du,23.134900\du){$down.\mathbf{0}$};
\definecolor{dialinecolor}{rgb}{0.000000, 0.000000, 0.000000}
\pgfsetstrokecolor{dialinecolor}
\node[anchor=west] at (14.052100\du,12.434900\du){$up$};
\pgfsetlinewidth{0.300000\du}
\pgfsetdash{}{0pt}
\pgfsetdash{}{0pt}
\pgfsetbuttcap
{
\definecolor{dialinecolor}{rgb}{0.000000, 0.000000, 0.000000}
\pgfsetfillcolor{dialinecolor}
\pgfsetarrowsend{stealth}
\definecolor{dialinecolor}{rgb}{0.000000, 0.000000, 0.000000}
\pgfsetstrokecolor{dialinecolor}
\draw (23.382936\du,17.198024\du)--(30.344564\du,20.439276\du);
}
\pgfsetlinewidth{0.300000\du}
\pgfsetdash{}{0pt}
\pgfsetdash{}{0pt}
\pgfsetbuttcap
{
\definecolor{dialinecolor}{rgb}{0.000000, 0.000000, 0.000000}
\pgfsetfillcolor{dialinecolor}
\pgfsetarrowsend{stealth}
\definecolor{dialinecolor}{rgb}{0.000000, 0.000000, 0.000000}
\pgfsetstrokecolor{dialinecolor}
\draw (30.435274\du,11.185514\du)--(23.403926\du,14.422786\du);
}
\pgfsetlinewidth{0.300000\du}
\pgfsetdash{}{0pt}
\pgfsetdash{}{0pt}
\pgfsetmiterjoin
\pgfsetbuttcap
{
\definecolor{dialinecolor}{rgb}{0.000000, 0.000000, 0.000000}
\pgfsetfillcolor{dialinecolor}
\pgfsetarrowsend{stealth}
\definecolor{dialinecolor}{rgb}{0.000000, 0.000000, 0.000000}
\pgfsetstrokecolor{dialinecolor}
\pgfpathmoveto{\pgfpoint{20.915707\du}{13.590197\du}}
\pgfpathcurveto{\pgfpoint{21.819507\du}{9.732697\du}}{\pgfpoint{25.759551\du}{7.976386\du}}{\pgfpoint{29.993151\du}{8.448806\du}}
\pgfusepath{stroke}
}
\pgfsetlinewidth{0.300000\du}
\pgfsetdash{}{0pt}
\pgfsetdash{}{0pt}
\pgfsetbuttcap
\pgfsetmiterjoin
\pgfsetlinewidth{0.300000\du}
\pgfsetbuttcap
\pgfsetmiterjoin
\pgfsetdash{}{0pt}
\definecolor{dialinecolor}{rgb}{1.000000, 1.000000, 1.000000}
\pgfsetfillcolor{dialinecolor}
\pgfpathellipse{\pgfpoint{35.133617\du}{9.022417\du}}{\pgfpoint{3.522418\du}{0\du}}{\pgfpoint{0\du}{3.522418\du}}
\pgfusepath{fill}
\definecolor{dialinecolor}{rgb}{0.000000, 0.000000, 0.000000}
\pgfsetstrokecolor{dialinecolor}
\pgfpathellipse{\pgfpoint{35.133617\du}{9.022417\du}}{\pgfpoint{3.522418\du}{0\du}}{\pgfpoint{0\du}{3.522418\du}}
\pgfusepath{stroke}
\pgfsetlinewidth{0.030000\du}
\pgfsetbuttcap
\pgfsetmiterjoin
\pgfsetdash{}{0pt}
\definecolor{dialinecolor}{rgb}{0.000000, 0.000000, 0.000000}
\pgfsetstrokecolor{dialinecolor}
\pgfpathellipse{\pgfpoint{35.133617\du}{9.022417\du}}{\pgfpoint{3.522418\du}{0\du}}{\pgfpoint{0\du}{3.522418\du}}
\pgfusepath{stroke}
\pgfsetlinewidth{0.100000\du}
\pgfsetdash{}{0pt}
\pgfsetdash{}{0pt}
\pgfsetbuttcap
\pgfsetmiterjoin
\pgfsetlinewidth{0.100000\du}
\pgfsetbuttcap
\pgfsetmiterjoin
\pgfsetdash{}{0pt}
\definecolor{dialinecolor}{rgb}{0.000000, 0.000000, 0.000000}
\pgfsetfillcolor{dialinecolor}
\pgfpathellipse{\pgfpoint{35.141115\du}{8.929915\du}}{\pgfpoint{0.904915\du}{0\du}}{\pgfpoint{0\du}{0.904915\du}}
\pgfusepath{fill}
\definecolor{dialinecolor}{rgb}{0.000000, 0.000000, 0.000000}
\pgfsetstrokecolor{dialinecolor}
\pgfpathellipse{\pgfpoint{35.141115\du}{8.929915\du}}{\pgfpoint{0.904915\du}{0\du}}{\pgfpoint{0\du}{0.904915\du}}
\pgfusepath{stroke}
\pgfsetlinewidth{0.010000\du}
\pgfsetbuttcap
\pgfsetmiterjoin
\pgfsetdash{}{0pt}
\definecolor{dialinecolor}{rgb}{0.000000, 0.000000, 0.000000}
\pgfsetstrokecolor{dialinecolor}
\pgfpathellipse{\pgfpoint{35.141115\du}{8.929915\du}}{\pgfpoint{0.904915\du}{0\du}}{\pgfpoint{0\du}{0.904915\du}}
\pgfusepath{stroke}
\pgfsetlinewidth{0.300000\du}
\pgfsetdash{}{0pt}
\pgfsetdash{}{0pt}
\pgfsetmiterjoin
{\pgfsetcornersarced{\pgfpoint{0.000000\du}{0.000000\du}}\definecolor{dialinecolor}{rgb}{0.000000, 0.000000, 0.000000}
\pgfsetfillcolor{dialinecolor}
\fill (15.111200\du,15.200000\du)--(15.111200\du,16.415000\du)--(25.681200\du,16.415000\du)--(25.681200\du,15.200000\du)--cycle;
}{\pgfsetcornersarced{\pgfpoint{0.000000\du}{0.000000\du}}\definecolor{dialinecolor}{rgb}{0.000000, 0.000000, 0.000000}
\pgfsetstrokecolor{dialinecolor}
\draw (15.111200\du,15.200000\du)--(15.111200\du,16.415000\du)--(25.681200\du,16.415000\du)--(25.681200\du,15.200000\du)--cycle;
}\pgfsetlinewidth{0.300000\du}
\pgfsetdash{}{0pt}
\pgfsetdash{}{0pt}
\pgfsetbuttcap
\pgfsetmiterjoin
\pgfsetlinewidth{0.300000\du}
\pgfsetbuttcap
\pgfsetmiterjoin
\pgfsetdash{}{0pt}
\definecolor{dialinecolor}{rgb}{1.000000, 1.000000, 1.000000}
\pgfsetfillcolor{dialinecolor}
\pgfpathellipse{\pgfpoint{35.033617\du}{22.622418\du}}{\pgfpoint{3.522418\du}{0\du}}{\pgfpoint{0\du}{3.522418\du}}
\pgfusepath{fill}
\definecolor{dialinecolor}{rgb}{0.000000, 0.000000, 0.000000}
\pgfsetstrokecolor{dialinecolor}
\pgfpathellipse{\pgfpoint{35.033617\du}{22.622418\du}}{\pgfpoint{3.522418\du}{0\du}}{\pgfpoint{0\du}{3.522418\du}}
\pgfusepath{stroke}
\pgfsetlinewidth{0.030000\du}
\pgfsetbuttcap
\pgfsetmiterjoin
\pgfsetdash{}{0pt}
\definecolor{dialinecolor}{rgb}{0.000000, 0.000000, 0.000000}
\pgfsetstrokecolor{dialinecolor}
\pgfpathellipse{\pgfpoint{35.033617\du}{22.622418\du}}{\pgfpoint{3.522418\du}{0\du}}{\pgfpoint{0\du}{3.522418\du}}
\pgfusepath{stroke}
\pgfsetlinewidth{0.300000\du}
\pgfsetdash{}{0pt}
\pgfsetdash{}{0pt}
\pgfsetbuttcap
{
\definecolor{dialinecolor}{rgb}{0.000000, 0.000000, 0.000000}
\pgfsetfillcolor{dialinecolor}
\pgfsetarrowsend{stealth}
\definecolor{dialinecolor}{rgb}{0.000000, 0.000000, 0.000000}
\pgfsetstrokecolor{dialinecolor}
\draw (30.276501\du,24.652718\du)--(22.699099\du,27.886782\du);
}
\pgfsetlinewidth{0.300000\du}
\pgfsetdash{}{0pt}
\pgfsetdash{}{0pt}
\pgfsetmiterjoin
{\pgfsetcornersarced{\pgfpoint{0.000000\du}{0.000000\du}}\definecolor{dialinecolor}{rgb}{0.000000, 0.000000, 0.000000}
\pgfsetfillcolor{dialinecolor}
\fill (14.261200\du,28.625000\du)--(14.261200\du,29.840000\du)--(24.831200\du,29.840000\du)--(24.831200\du,28.625000\du)--cycle;
}{\pgfsetcornersarced{\pgfpoint{0.000000\du}{0.000000\du}}\definecolor{dialinecolor}{rgb}{0.000000, 0.000000, 0.000000}
\pgfsetstrokecolor{dialinecolor}
\draw (14.261200\du,28.625000\du)--(14.261200\du,29.840000\du)--(24.831200\du,29.840000\du)--(24.831200\du,28.625000\du)--cycle;
}
\definecolor{dialinecolor}{rgb}{0.000000, 0.000000, 0.000000}
\pgfsetstrokecolor{dialinecolor}
\node[anchor=west] at (11.505000\du,25.767500\du){$down$};
\end{tikzpicture}
\caption{The P/T system for a semi-counter.}
\label{semi-counter}
\end{figure}

\begin{example} \label{ex-semi-counter}{\bf (Semi-counter)}
A semi-counter process, i.e., a counter that cannot test for zero, can be described  by
the infinite-state process $A \eqdef $ $up. (down.\nil \para A)$. Observe that $\dec(A) =$ $ \{ up. (down.\nil \para A) \}$.
The only enabled transition is $\dec(A) \deriv{up} down.\nil \oplus  up.(down.\nil \para A)$.
Then, also transition $down.\nil \deriv{down} \emptyset$ is derivable. The finite P/T net $Net(A)$ is reported 
in Figure \ref{semi-counter}.
\end{example}

\begin{figure}[!ht]
\centering
\input{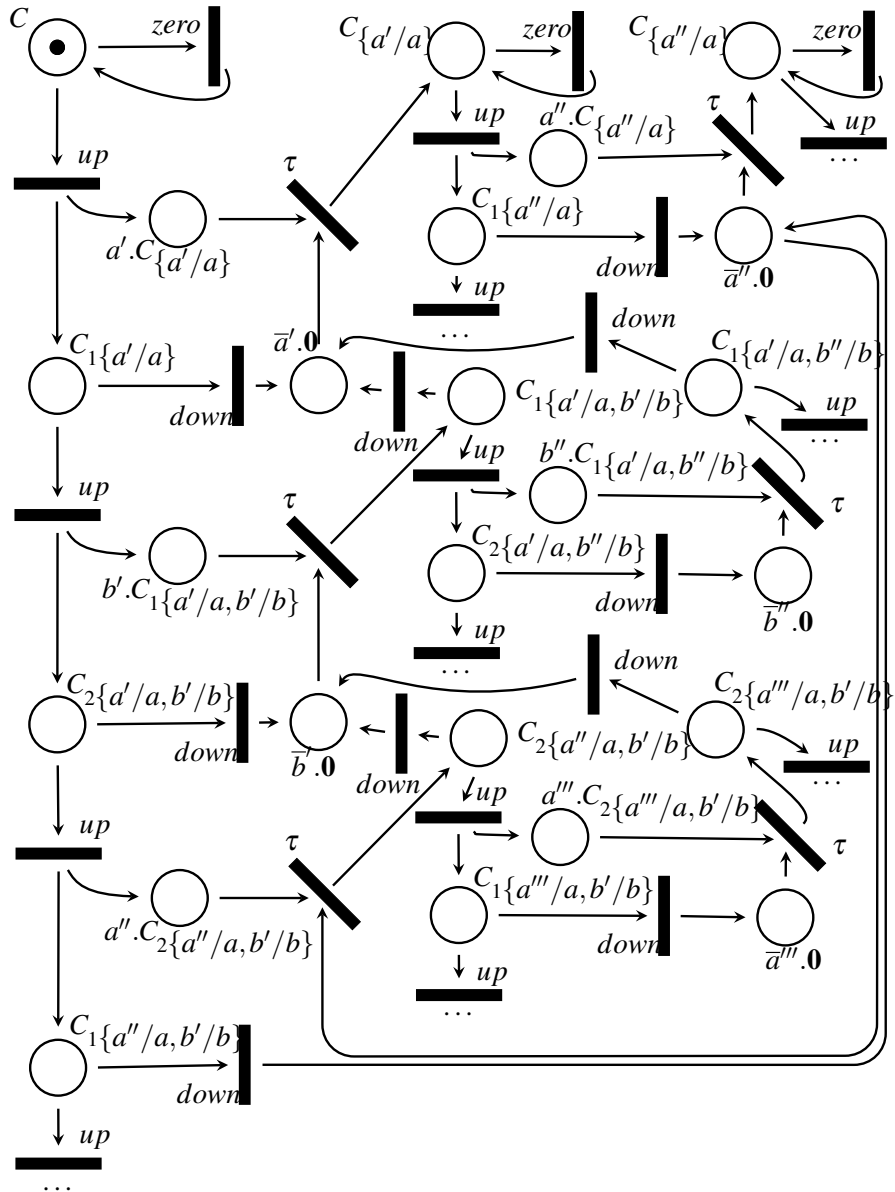}
\caption{The initial fragment of the P/T system for counter $C$.}
\label{counter-net}
\end{figure}

\begin{example}\label{counter}{\bf (Counter with test for zero)}
As an example of a CCS process that cannot be modeled by a finite P/T net, consider the following
specification of a (real) counter, as given in \cite{Tau}. 

$
\begin{array}{lcl}
C &  \eqdef &  zero.C + up.(\restr{a}(C_1 \para a.C))\\
C_1 &  \eqdef & down.\bar{a}.\nil  + up.(\restr{b}(C_2 \para b.C_1)))\\
C_2 &  \eqdef & (down.\bar{b}.\nil  + up.(\restr{a}(C_1 \para a.C_2)))\\
\end{array}
$

An initial fragment of the infinite P/T net $Net(C)$ is reported\footnote{For brevity,
we associate to a place the name of a constant instead of its definition, e.g. place $C$ should be called
$zero.C + up.(\restr{a}(C_1 \para a.C))$.} in Figure \ref{counter-net},
where successive unfoldings are due to syntactic substitutions applied to constants that generate new places.
Note also the peculiar way substitution is applied to restricted terms. 
\end{example}

\begin{figure}
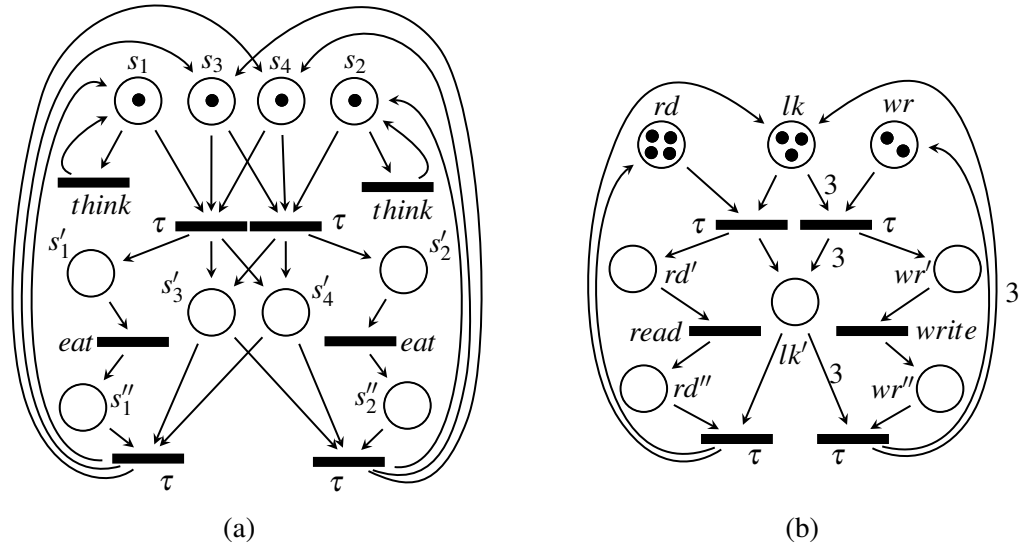

\centering
\begin{tabular}{ccc}
\input{diningphil-bg-net-tex} & 
&
\input{readerswriters-bg-red-tex} \\
(a) & & (b) 
\end{tabular}
\caption{(a) The net for two dining philosophers. (b) The net for concurrent readers/writers.}
\label{dining1-net}
\end{figure}

\begin{example} {\bf (Dining Philosophers)}
Consider the system $DF$ of Example \ref{ex-dining}. The marking $\dec(DF)$ is composed of the four 
places\footnote{Again, for brevity,
we associate to a place the name of a constant instead of its definition, e.g. $s_1 = phil_0$ while it should be 
$s_1 = think.phil_0 + \underline{up_0}.up_{1}.eat.\underline{dn_0}.dn_{1}.phil_0$.}
$s_1 = phil_0$, $s_2 =  phil_1$, $s_3 = fork_0$ and  $s_4 = fork_1$.
Initially, the two philosophers can think on their own:

$s_1  \deriv{think} s_1$ and  $s_2  \deriv{think} s_2$ 

\noindent
or can compete for the acquisition of  the two forks: 

$s_1 \oplus s_3 \oplus s_4 \deriv{\tau} s'_1 \oplus s'_3 \oplus s'_4$ and 

$s_2 \oplus s_3 \oplus s_4 \deriv{\tau} s'_2 \oplus s'_3 \oplus s'_4$ 

\noindent
where $s'_1 =  phil'_0$, $s'_2 = phil'_1$,
$s'_3 = \overline{down_0}.fork_0$, $s'_4 = \overline{down_1}.fork_1$\\
with, for $i = 0,1$,  $phil'_i = eat.\underline{down_i}.down_{i+1(mod 2)}.phil_i$.
Now two further alternative transitions are derivable, namely:

$s'_1  \deriv{eat} s''_1$ and  $s'_2  \deriv{eat} s''_2$ 

\noindent
where  $s''_1 =  phil''_0$, $s''_2 = phil''_1$,
with, for $i = 0,1$,  $phil''_i = \underline{down_i}.down_{i+1(mod 2)}.phil_i$. Finally,

$s''_1 \oplus s'_3 \oplus s'_4 \deriv{\tau} s_1 \oplus s_3 \oplus s_4$ and 

$s''_2 \oplus s'_3 \oplus s'_4 \deriv{\tau} s_2 \oplus s_3 \oplus s_4$

\noindent
and we are back to the initial marking $\dec(DF)$. The resulting $Net(DF)$ is reported in Figure \ref{dining1-net}(a).
Note that the two philosophers can never eat at the same time, i.e., in no reachable marking $m$
we have that  $m(s'_1) = 1 = m(s'_2)$.\\[-.4cm]
\end{example}

\begin{example}{\bf (Concurrent readers and writers)}
Let us consider $Sys$ of Example \ref{ex-conc-rw}. The multiset $\dec(Sys)$ is $4 rd
\; \oplus \; 3 lk \; \oplus \;  2 wr
\; \oplus \restr{l} \; \oplus \; \restr{u}$, where $rd = l.read.u.R$, $lk = \overline{l}.\overline{u}.L$
and $wr = \underline{l}.\underline{l}.l.write.$ $\underline{u}.\underline{u}.u.W$.
One of the two possible initial transitions is $wr \oplus 3lk \deriv{\tau} wr' \oplus 3lk'$, where $wr' = 
write.\underline{u}.\underline{u}.u.W$
and $lk' = \overline{u}.L$. After such a transition, no reader can read, as all the locks are busy.
The other possible initial transition is $rd \oplus lk \deriv{\tau} rd' \oplus lk'$, where $rd' = read.u.R$. 
The resulting P/T net $Net(Sys)$ is depicted in Figure \ref{dining1-net}(b).\\[-.6cm]
\end{example}

\section{Properties of the net semantics\\[-.6cm]}\label{propBG}

In this section, we present some results about the net semantics we have defined. First we give 
a soundness result, 
namely that the interleaving marking graph associated to $Net(p)$ for any Multi-CCS term $p$ is bisimilar to its
transition system.
Then we discuss finiteness conditions on the net semantics. In particular, we
single out a subclass of  Multi-CCS processes whose semantics always generates finite P/T nets.
This subclass, we call {\em finite-net} processes, is rather rich, as the parallel operator is allowed to occur inside 
the body of recursively defined constants. Hence, finite-net processes may be infinite-state processes,
(i.e., the associated labeled transition 
system may contain infinitely many states), as illustrated in Example \ref{ex-semi-counter}. 

\subsection{Soundness}

\begin{proposition}\label{unavia}
For any process $p \in {\mathcal P}$,  if $p \deriv{\sigma} p'$ then there exists $t \in T_p$
such that $dec(p)[t\rangle K$ with $l(t) = \sigma$ and $K \sim \dec(p')$.

\proof By induction on the proof of $p \deriv{\sigma} p'$. 
\fine
\end{proposition}

\begin{proposition}\label{altravia}
For any process $p \in {\mathcal P}$, if there exists $t \in T_p$ such that $dec(p)[t\rangle K$ with $l(t) = \sigma$, 
then there exists $p'$ such that $p \deriv{\sigma} p'$ and $K = \dec(p')$.

\proof
By induction on (the definition of) $\dec(p)$ and then by induction 
on the proof of  $t$.
\fine
\end{proposition}

\begin{theorem}
For any process $p \in {\mathcal P}$, $p \sim \dec(p)$.

\proof
Relation $R = \{ (p, \dec(q)) \mid  p, q \in {\mathcal P}, \, \; \dec(p) \sim \dec(q)\}$  is a bisimulation, due to Proposition \ref{unavia} 
(together with Proposition \ref{decprop}) and Proposition \ref{altravia}.
\fine
\end{theorem}

\subsection{Finiteness} \label{opt}

The net semantics often generates finite nets. However, 
the generation of an infinite system may be due to one of the following three facts.
First, the decomposition rule for restriction requires the
generation of a fresh name; hence, if this operator lies inside a 
recursive definition, an infinite set of fresh names (i.e., of places) may be required.
Second, we have to impose a finite bound to the number of constants that can be used in
a process definition. E.g., process $b.A_0$, with the family of process constants 
$A_i \eqdef a_i.A_{i+1}$ for $i \in \nat$, is not allowed.
Third, as the synchronization relation is too generous (it may produce
infinitely many transitions even for a net with finitely many places, as the following example shows),
we have to impose a restriction over $Sync$, that disables transactional communication but allows for
multi-party synchronization.

\begin{example}
Consider $B \eqdef \underline{a}.\bar a.(B \para B)$. $Net(B)$ has just one place $p = \underline{a}.\bar a.(B \para B)$,
but infinitely many transitions! The only possible initial net transition
is $p \deriv{a \bar a} 2p$.
Now transition $2p \deriv{a \bar a} 4p$ is possible, and then $4p \deriv{a \bar a} 8p$, and so on ad infinitum. 
\end{example}

\begin{definition} \label{deffnmccs}
The  {\em finite-net} Multi-CCS processes are the processes generated by the following syntax\\[-.7cm]
\begin{eqnarray*}
s & ::= & \nil \mid \mu.t \mid \underline\mu.t \mid s+s \\\\[-.6cm]
t & ::= & s \mid t \para t \mid  C \\\\[-.6cm]
p & ::= &   t  \mid \restr{a}p \mid p \para p\\[-.6cm]
\end{eqnarray*}
where a constant $C$ has associated a term of type $t$, i.e., $C \eqdef t$ and the number of constants
involved in any process definition is always finite.

The semantics of finite-net Multi-CCS is the same as provided for Multi-CCS in Tables~\ref{rules} and \ref{sync}, 
with the following additional constraint on rule $(Com)$: $Sync(\sigma_1, \sigma_2, \sigma)$
is applicable only if $|\sigma_1| = 1$ or $|\sigma_2| = 1$.
\fine 
\end{definition}

\begin{theorem} \label{finitenet}
Let $p$ be a finite-net process. Then the subnet $Net(p)$ associated to $p$ is finite.
\fine
\end{theorem}

\section{A process term for any finite P/T net}

Now  the converse problem: given a finite P/T system $N(m_0)$, can we single out a finite-net 
process $p_{N(m_0)}$ such that $Cl(p_{N(m_0)})$ and $N(m_0)$ are isomorphic? 
The answer is positive, hence providing a language for finite P/T Petri nets. 

The translation from nets to processes
we present 
takes a restricted name $y_i$ for any place $s_i$; this is used to distinguish syntactically all the places,
so that no fusion is possible when applying the reduced net reverse translation.
Moreover, it considers a restricted name $x_j$ for each transition $t_j$, that is used to synchronize all the components
that participate in $t_j$. The constant $C_i$ associated to a place $s_i$ has a summand for each transition
which $s_i$ is in the preset of. Among the many places in the preset of $t_j$, the one connected with an arc of 
minimal weight
(and if more than one is so, then the one with minimal index) plays the role of {\em leader} of the multiparty 
synchronization (i.e., the process performing the atomic sequence of inputs $x_j$ to be synchronized with single outputs
$\bar x_j$ performed by the other participants).

\begin{definition}\label{translate}
Let $N(m_0) = (S, A, T, m_0)$, with $S = \{s_1, \ldots, s_n\}$ and $T = \{t_1, \ldots, t_v\}$. 
Function \linebreak $\inet{N(m_0)}$ from finite P/T systems to finite-net processes is defined as (for fresh $x_i$ and $y_j$)\\[-.6cm]
\begin{eqnarray*}
\inet{N(m_0)} = & \restr{x_1\ldots x_v}  \restr{y_1\ldots y_n}
                    \; ( \underbrace{ C_1 | \cdots | C_1}_{m_0(s_1)} | \cdots | 
                    \underbrace{ C_n | \cdots | C_n}_{m_0(s_n)} )\\[-.9cm]
\end{eqnarray*}
where each $C_i$ has a defining equation\\[-.4cm]
\[ C_i \eqdef{} c_i^1 + \cdots + c_i^{p_i}  + y_i.\nil\]\\[-.4cm]
where $p_i$ is the size of  $\post{s_i} = \{t_{i_1}, \ldots, t_{i_{p_i}} \} \subseteq T$ such that $ s_i \in dom(\pre t)$ for each $t \in \post{s_i}$.
Let $d_{i_j} = \sum_k \big( \pre t_{i_j} (s_k) \big) - 1$ and $a_{i_j} = l(t_{i_j})$. Then, each $c_i^j$ is equal to\\[-.5cm]
\begin{itemize}
\item $a_{i_j}.\Pi_{i_j}$ 
              if $d_{i_j} = 0$ (no synchronization as $\pre t _{i_j} = s_i$);
\item $\overline{x}_{i_j}.\nil$ 
              if the previous condition does not hold, and
              $\pre t_{i_j} (s_i) > \pre t_{i_j} (s_{i'})$ for some $i'$ or 
              $\pre t_{i_j} (s_i) = \pre t_{i_j} (s_{i'})$ for some $i' < i$
             (i.e., $s_i$ is not the leader for the synchronization on $t_{i_j}$)
\item $\underbrace{ \underline{x}_{i_j}. \cdots . \underline{x}_{i_j}}_{d_{i_j}} . a_{i_j}.\Pi_{i_j}$
              if the previous conditions do not hold (i.e., $s_i$ is the \\[-3.5ex] 
              leader), and
              $\pre t_{i_j} (s_i) = 1$; if $a_{i_j} = \tau$, $c_i^j$ is simplified to
              $\overbrace{ \underline{x}_{i_j}. \cdots . \underline{x}_{i_j}}^{d_{i_j} - 1} . x_{i_j}.\Pi_{i_j}$;
\item $\overline{x}_{i_j}.\nil + \underbrace{ \underline{x}_{i_j}. \cdots . \underline{x}_{i_j}}_{d_{i_j} } . a_{i_j}.\Pi_{i_j}$
              otherwise (i.e., $s_i$ is the leader and the arc has weight  $> 1$).
\end{itemize}
Finally, each $\Pi_{i_j}$ is defined as
$ \Pi_{i_j} =  \underbrace{ C_1 | \cdots | C_1}_{ \post{t_{i_j}} (s_1) } | \cdots | 
                    \underbrace{ C_n | \cdots | C_n}_{ \post{t_{i_j}} (s_n) } . $
\end{definition}


\begin{paragraph}
{\bf Remark: (CCS nets)} 
Let us call {\em CCS nets} the class of P/T nets where transitions have only one input arc (with weight 1) or 
two input arcs (with weight 1) but labelled by $\tau$. It is not difficult to see that, given a CCS net $N(m_0)$
the resulting process term $\inet{N(m_0)}$ is a finite-net CCS terms (i.e., a term without strong prefixing). 
\end{paragraph}

\begin{figure}
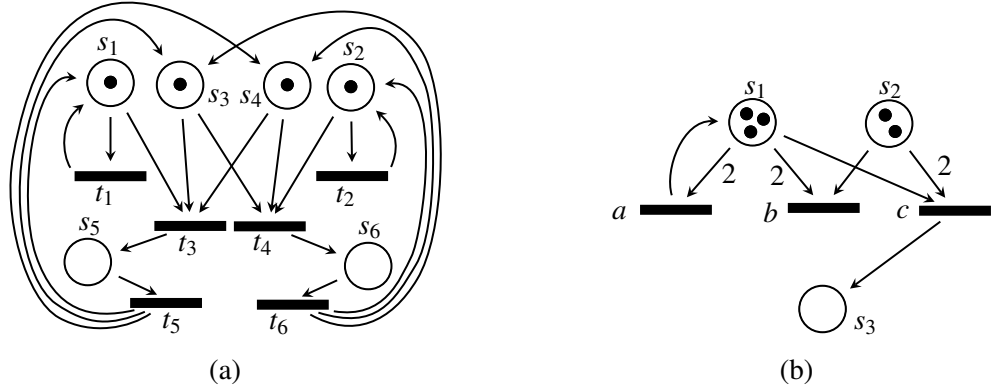

\centering

\begin{tabular}{ccc}
\input{diningphil2-bg-red-anonim-tex} & \hspace{1cm} & 
\input{altroesempio-tex} \\
(a) & & (b) 
\end{tabular}
\caption{(a) Alternative two philosophers' net. (b) A simple net}
\label{df2net}
\end{figure}

\begin{example}
Consider the net $N$ depicted in Figure \ref{df2net}(a), where we assume that $l(t_1) = l(t_2) = think$,
$l(t_3) =$  $ l(t_4) = \tau$ and $l(t_5) = l(t_6) = eat$.
Clearly, it is a different solution to the dining
philosophers problem, where forks (places $s_3$ and $s_4$) are resources that are consumed and then regenerated. 
Applying the translation above, we obtain the finite-net process
$\inet{N(m_0)} =  \restr{x_1\ldots x_6}  \restr{y_1\ldots y_6} ( C_1\para  C_2 \para C_3 \para C_4)$
where 

$
\begin{array}{lcllcl}
C_1 & \eqdef & think.C_1 + \underline{x_3}.x_3.C_5 + y_1.\nil & \; \; 
C_2 & \eqdef & think.C_2 + \underline{x_4}.x_4.C_6 + y_2.\nil\\
C_3 & \eqdef & \overline{x_3}.\nil + \overline{x_4}.\nil + y_3.\nil & \; \;
C_4 & \eqdef & \overline{x_3}.\nil + \overline{x_4}.\nil + y_4.\nil \\
C_5 & \eqdef &eat.(C_1 \para C_3 \para C_4) + y_5.\nil & \; \;
C_6 & \eqdef &eat.(C_2 \para C_3 \para C_4) + y_6.\nil\\
\end{array}
$

\noindent
Note that $C_3$ and $C_4$ differ for the last summand only. If the restricted names $y_3$ and $y_4$
were omitted, $Net(\inet{N(m_0)})$ would be a different net where places $s_3$ and $s_4$ are fused
in a new place with two tokens.
 
$\inet{N(m_0)}$ generates an infinite-state labeled transition system, because of the nesting of parallel 
operator inside recursively defined constants. 
However, its behavior is actually finite: indeed, it generates a finite safe P/T net, hence with a finite interleaving marking 
graph, which is bisimilar to its infinite-state labeled transition system. 
\end{example}

\begin{example}
Consider the net $N(m_0)$ of Figure \ref{df2net}(b).
Applying the translation above, we obtain the finite-net process
$\inet{N(m_0)} =  \restr{x_1 x_2  x_3}$ $  \restr{y_1 y_2  y_3} $ $( C_1 \para C_1 \para C_1 \para  C_2 \para C_2)$
where 

\noindent
$
C_1 \;  \eqdef \;  \overline{x_1}.\nil  +   \underline{x_1}.a.C_1 +   \overline{x_2}.\nil   + \underline{x_3}.\underline{x_3}.c.C_3 + y_1.\nil\\
C_2 \; \eqdef  \;   \underline{x_2}.\underline{x_2}.b.\nil +  \overline{x_3}.\nil +  y_2.\nil \; \; \; \; \hspace{1cm}
C_3 \;  \eqdef \;  y_3.\nil 
$
\end{example}


\begin{theorem}\label{represent}
Let $N(m_0)$ be a finite reduced system. Then, $Net(\inet{N(m_0)})$ is isomorphic to $N(m_0)$.
%
\fine
\end{theorem}

\begin{corollary}
Let $N(m_0)$ be a finite reduced CCS net. Then, $ \inet{N(m_0)}$ is a CCS process term and
$Net(\inet{N(m_0)})$ is isomorphic to $N(m_0)$.
\fine
\end{corollary}

\section{Conclusion}

The class of finite-net Multi-CCS processes represents a language for describing
finite P/T nets. This is not the only language expressing P/T nets: the first (and only other) one is
Mayr's PRS \cite{Mayr}, which however is rather far from a typical process algebra as its basic building blocks 
are rewrite rules (instead of actions) and, for instance,
it does not contain any scope operator like restriction or hiding. We think 
the language we have identified can be used
in order to cross-fertilize the areas of process calculi and Petri nets.
In  one direction, it opens, e.g., the problem of
finding axiomatizations of Petri nets behaviours. For instance,
net isomorphism induces a lot of equations over Multi-CCS
terms. Just to mention a few, parallel composition is associative, commutative with
$\nil$ as neutral element,  terms that differ only for alpha-conversion of bound names
are identified, 
the sum operator  is associative, commutative and, if the sequential term $p$ is not
$\nil$, then also $p + \nil = p$ and $p + p = p$ hold.
Even if the problem of finding a complete set of axioms characterizing net
isomorphism is probably out-of-reach, nonetheless, the axioms we have identified are interesting as they
include those forming the structural congruence
for CCS \cite{Mil99}, hence validating their use.
On the other direction, Petri net theory can offer a lot of support to
process algebra. Some useful properties are decidable
for finite P/T nets (e.g., reachability, liveness, coverability -- see e.g., \cite{Reisig} --
model-checking of linear time $\mu$-calculus formulae \cite{Esp94}) and so
also the (infinite-state systems of) finite-net Multi-CCS processes can
be checked against these properties. Moreover, P/T nets are equipped with
non-interleaving semantics, where
parallel composition is not reduced to sum and prefixing, and these
semantics can be used fruitfully to check
causality-based properties, useful, e.g., in error recovery.

As a final remark, we want to stress that our net semantics is the first
one based on unsafe labeled P/T nets for a rich process algebra including CCS as a subcalculus.
Indeed, our net semantics improves over previous work. 
Goltz's result \cite{Ulla1,Ulla} are limited to CCS without restriction; we define our net semantics in a different style
(operational) and additionally we cope with restriction and strong prefixing.
Degano, De Nicola, Montanari \cite{DDMa} and Olderog's approach \cite{Old}
is somehow complementary in style, as it builds directly over
the SOS semantics of CCS. Their construction generates {\em safe} P/T
nets which are finite only for regular CCS processes (i.e., processes where
restriction and parallel composition cannot
occur inside recursion). Moreover, this approach has never been applied to a 
process algebra whose labeled operational semantics is defined modulo a 
structural congruence.
Similar concerns are for PBC \cite{BDK}, whose semantics is given in terms
of safe P/T nets. Nonetheless, PBC can express  ``programmable'' multiway synchronization by means
of its relabeling operators (somehow similar to Multi-CCS), and so, in principle, if equipped with an unsafe semantics
it might also serve as a language expressing general P/T nets. On the contrary, we conjecture
that it is not possible to obtain a representation theorem such as Theorem \ref{represent} 
based on CSP \cite{Hoare}. 

Our work is somehow indebted to the earlier work of Busi \& Gorrieri \cite{BG09} on giving labeled net semantics
to $\pi$-calculus in terms of P/T nets with inhibitor arcs; our solution simplifies this approach for CCS and Multi-CCS
because we do not need inhibitors. In particular, already in that paper it is observed that finite-net $\pi$-calculus
processes originate finite net P/T net systems (with inhibitor arcs). 
Similar observations on the interplay between parallel composition and restriction in recursive definitions, in
 different contexts, has been done also by others, e.g., \cite{Aranda}.
Also important is the work of Meyer \cite{Mey,MG09} in providing an unlabeled
P/T net semantics for a fragment of $\pi$-calculus; the main difference is that his semantics may offer a finite
net representation also for some processes where restriction occurs inside recursion, but the price to pay is that
the resulting net semantics may be not correct from a causality point of view. We conjecture that his technique is 
not applicable to Multi-CCS.

Future work will be devoted to define compositional (denotational in style) unsafe net semantics for Multi-CCS,
generalizing work of Goltz \cite{Ulla1} and Taubner \cite{Tau}. \\[-.6cm]

\section*{Acknowledgment\\[-.8cm]}

The first author would like to thank Eike Best, Philippe Darondeau and Pierpaolo Degano and the anonymous referees
for helpful comments.\\[-.6cm]

\end{document}